\documentclass[journal,draftcls,onecolumn]{IEEEtran}
\usepackage{transparent}
\usepackage[english]{babel}
\usepackage{float}
\usepackage[utf8]{inputenc}
\usepackage{amsmath}
\usepackage{enumitem}
\usepackage{graphicx}
\usepackage{clipboard}
\usepackage[usenames, dvipsnames]{color}
\usepackage{amssymb}
\usepackage{xfrac}
\usepackage{amssymb}
\usepackage{amsthm}
\usepackage{verbatim}
\usepackage{url}
\usepackage{cite}
\usepackage{epstopdf}
\usepackage{bbm}
 \usepackage{lettrine}
\usepackage{xcolor}
\usepackage{mathtools}
\usepackage{xr}

\newcommand*{\QEDB}{\hfill\ensuremath{\square}}%

\graphicspath{{Figures/}}

\newcommand{\btau}{\boldsymbol{\tau}}
\newcommand{\bd}{\boldsymbol}
\newcommand{\T}{\mathcal{T}}
\newtheorem{defn}{Definition}
\newtheorem{lem}{Lemma}
\newtheorem{prop}{Proposition}
\newtheorem{thm}{Theorem}
\newtheorem{ex}{Example}

\newtheorem*{rmk}{Remark}

\DeclareMathOperator*{\argmax}{arg\,max}
\DeclareMathOperator*{\argmin}{arg\,min}

\newclipboard{myclipboard}

\ifCLASSINFOpdf

\else

\fi

\begin{document}

\title{Participation of an Energy Storage Aggregator in Electricity Markets}

\author{Jesus E. Contreras-Oca\~na,~\IEEEmembership{Student Member,~IEEE,}
        Miguel A. Ortega-Vazquez,~\IEEEmembership{Senior Member,~IEEE,}\\
        Baosen Zhang,~\IEEEmembership{Member,~IEEE}
\thanks{The authors are with the Department of Electrical Engineering at the University of Washington. Emails: \{jcontrer, maov, zhangbao\}@uw.edu. This work is partially supported by the University of Washington Clean Energy Institute.}
}

\maketitle

\begin{abstract}
An important function of aggregators is to enable the participation of small energy storage units in electricity markets. This paper studies two generally overlooked aspects related to aggregators of energy storage: i) the relationship between the aggregator and its constituent storage units and ii) the aggregator's effect on system welfare.  Regarding i), we show that short-term outcomes can be Pareto-inefficient: all players could be better-off.  In practice, however, aggregators and storage units are likely to engage in long rather than short-term relationships. Using Nash Bargaining Theory, we show that aggregators and storage units are likely to cooperate in the long-term. A rigorous understanding of the aggregator-storage unit relationship is fundamental to model the aggregator's participation in the market.  Regarding ii), we first show that a profit-seeking energy storage aggregator is always beneficial to the system when compared to a system without storage, regardless of size or market power the aggregator may have. However, due to market power, a monopolist aggregator may act in a socially suboptimal manner. We propose a pricing scheme designed to mitigate market power abuse by the aggregator. This pricing scheme has several important characteristics: its formulation requires no private information, it incentivizes a rational aggregator to behave in a socially optimal manner, and allows for regulation of the aggregator's profit.
\end{abstract}


\section{Introduction} \vspace{0pt}
\label{sec:intro}
\lettrine{T}{he} adoption of household-level energy storage systems is expected to increase rapidly in the coming years (residential energy storage grew by 405\% in 2015) and become a significant share of the total U.S. energy storage deployment~\cite{storage_report_2016}.  These storage units (SUs) have the potential of selling services to the power grid~\cite{DenholmEtAl2013} but may not be able to directly do so for two main reasons: i) their individual capacities are smaller than the required minimum~\cite{ISONE_energy_storage,PJM_DR}; and ii) the large number of SUs would make their management difficult even if they are allowed to participate~\cite{Sarker_Dvorkin_Ortega_2016}.  Therefore, \emph{aggregators} act as mediators between SUs and the power system~\cite{BurgerEtAl}.

A number of studies regarding the operation and market strategies of aggregators have been conducted. For instance, the authors of~\cite{Sarker_Dvorkin_Ortega_2016, MOV_B_S_2013} study the aggregation of a fleet of electric vehicles while~\cite{Chen_Wang_2014} studies the coordination of groups of residential consumers for demand response. These studies focus on the interactions between aggregator and grid, with little regard to the interactions between the aggregator and its constituents.  On a topic closely related to this work, authors of~\cite{Nguyen_etal_2015} use Nash Bargaining Theory to determine the compensation of reactive power providers.  These works, however, model aggregators as price-takers in the market.  This assumption could be unrealistic since aggregators of SUs can be both large and flexible, exactly the type of agents that have market power.  This paper bridges the gap between current studies and a realistic market setting by addressing two important aspects: i) How should the interactions between SUs and an aggregator be modeled when both are profit-seeking entities? ii) How should we quantify and mitigate the market power and price-anticipatory behavior\footnote{Price-anticipatory behavior arises when players are aware of their own market power and the effect of their actions on the market price.} of these aggregators?

Both of these questions have been studied, although normally not in an integral manner.  In~\cite{Momber_Wogrin_2016}, the authors focus on the distributed optimization problem of an electric vehicle (EV) aggregator assuming exogenous market prices.  The authors in~\cite{Weywadt2012,VayaEtAl2015} study the problem subjected to power flow constraints.  In~\cite{Wu_etal_2012}, the authors study the interactions of an EV aggregator and its EVs when providing frequency regulation.  However, these studies do not address how the total profit should be divided between an aggregator and its SUs.  The problem of profit allocation is not obvious since neither party can participate in the market alone: the aggregator has no physical assets and the individual SUs are too small to participate.  In this work, we use Nash Bargaining Theory to derive a model of the profit-split between the aggregator and SUs.

Other works study the impact of price-anticipatory behavior by aggregators. The works in~\cite{Wu,VagropoulosEtAl2013,RoteringEtAl2011} study strategic bidding of aggregators through optimization frameworks.  These studies provide understanding of the aggregator's operational problem but do not address the question of the effect of market power on the system welfare.  From general literature on markets and game theory, it is well-known that competition increases efficiency and reduces market power.  However, it is likely that only a small number of aggregators would be available for a customer to choose from.  In this work, we draw qualitative insights by considering a single aggregator responsible for a set of customers.  Then the natural question becomes: in a system with a small number of aggregators, would the market power of the aggregators outweigh the benefit of having storage in the system?

This paper addresses: i) the problem of profit allocation between the aggregator and its users, and ii) the question of impact of the aggregator on the system welfare.  To address i), we adopt a formal Nash Bargaining Process~\cite{Muthoo1999,Nash-1950} to model the negotiations between the aggregator and its users.  A key feature of the proposed model is that it captures the realistic setting where the parties interact on a continuous basis.   Then we establish the necessary profit splitting conditions such that the bargaining agreement can be sustained indefinitely.

\begin{figure}[tb]
	\centering
	\includegraphics[width=0.5\textwidth]{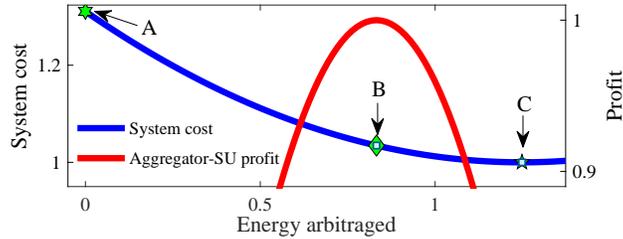} \vspace{0pt}
	\caption{ System cost~(blue curve) and the profit of a storage aggregator~(red curve). Without energy storage, the system operates are point A. With an aggregator that can exercise market power, the system operates at point B. If the storage units are fully controllable by a social planner, then the system cost is at point C. In this paper, we show that point B always costs less than point A, that is, having storage -- no matter the market power -- is beneficial to the system. We then present a way to mitigate the market power of the aggregator such that the system can operate at point C. The system and energy storage parameters used to compute these curves are given in Example~\ref{Ex:equential} and used throughout the rest of the examples.} \label{profit_and_SW}
\end{figure}

To answer ii), we adopt two perspectives.  First, we show that the participation of SUs \emph{always} benefits the system. Consider Fig.~\ref{profit_and_SW}: Point A represents the system cost without SUs while point B represents the system cost when the SUs are controlled by a price-anticipatory aggregator. While A is higher than B, the aggregator does not act in a \emph{socially optimal} fashion (due to its market power), limiting the maximum possible gain from having energy storage in the system.  We propose a pricing scheme that effectively mitigates the market power of aggregators, incentivizing them to operate at the social optimum (point C in Fig. \ref{profit_and_SW}). The only assumption we make on the aggregator is that it is profit-seeking and does not act in a purely adversarial manner.

\begin{figure}[tb]
	\centering
	\includegraphics[width=0.5\textwidth]{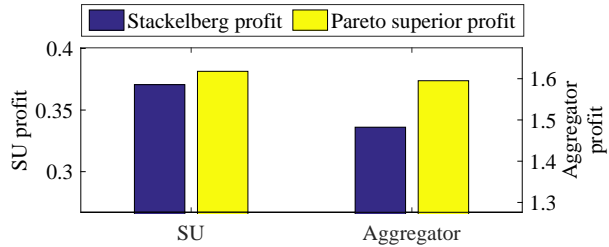} \vspace{-0pt}
	\caption{The blue bars are the respective profits for the SU and the aggregator under the parameters defined in Example \ref{Ex:equential}. These are inefficient in a Pareto sense since there exists an operating point where both parties receive higher profits~(yellow bars).}
	\vspace{-0pt}
	\label{fig:ex1}
\end{figure}

In summary, the contributions of this work are:
\begin{itemize}
	\item[i)] \emph{A model of the interactions between an aggregator, its managed SUs, and the electricity market.}  First, we introduce the model of short-term interactions between aggregator and SUs: the aggregator and SUs play a single-leader, multi-follower Stackelberg Game.  We show that the outcome of this non-cooperative game can be Pareto-inefficient~(see Fig.~\ref{fig:ex1}).  In reality, however, the aggregator and SUs are likely to interact repeatedly.  We demonstrate the existence of profit-splitting schemes in which cooperation by rational agents is stable in the long-term.  Finally, we use the Nash Bargaining Model to show that the aggregator and SUs split a Pareto-optimal profit in a way that ensures long-term cooperation.
	\item[ii)] \emph{A simplified model of a set of SUs managed by an aggregator.} The Pareto-optimal agreement allows us to replace a complicated system composed of many agents (\emph{i.e.}, the aggregator and SUs) with a simpler aggregate profit maximization problem. The benefits of this simplified model is two-fold. First, it provides and justifies a model of the aggregator's market participation strategy. Second, it provides a computationally tractable model of the actions of numerous SUs under an aggregator.
	\item[iii)] \emph{An assessment of the impact the aggregator in the system welfare.}  First we show that a profit-seeking aggregator is always beneficial to the system.  Then, we compare the socially optimal system cost with the system cost derived from the bargaining agreement.
	\item[iv)] Finally, we propose \emph{a market power-mitigating pricing scheme} that: 1) incentivizes self-interested aggregators to behave in a socially optimal fashion, 2) requires no private information for its formulation, and 3) allows for regulation of the aggregator's profit.
\end{itemize}

In this paper, we emphasize the theoretical foundations of our work and illustrate our propositions using small numerical examples. A larger case study, however, is useful to appreciate some of the nuances that would arise in more complicated systems. The interested reader is referred to~\cite{Contreras_PESGM2017} for a case study on the 24 bus IEEE Reliability Test System (RTS).

This paper is organized as follows. Section~\ref{section:Market} introduces the market, SU, and aggregator models. Section \ref{sec:agg} demonstrates that having a profit-seeking aggregator is always beneficial to the system. Section \ref{sec:seqgame} introduces the short-term interaction model between the aggregator and the SUs. Section~\ref{section:longtermcoop} introduces both the framework used to model long-term interactions and the Nash Bargaining Model. In section~\ref{sec:mitigation} we use the Nash Bargaining Solution to model how the aggregator bids into the market and propose a pricing scheme to mitigate its market power. Section \ref{sec:conclusion} concludes this paper.

\vspace{0pt}
\section{Models and Problem Setup} 
\begin{figure}[tb]
	\centering
	\includegraphics[width=0.3\textwidth]{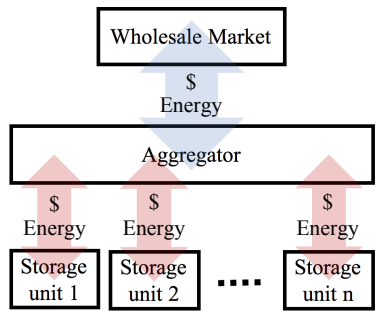}
	\caption{Market-aggregator-SUs framework.} \label{fig:Diagram}
\end{figure}

\vspace{-0pt}
\label{section:Market}
This section introduces the models used throughout the paper: the market structure, the SUs, and the aggregator. We assume that the SUs do not participate directly in the market. Instead, they interact with an aggregator which has access to the wholesale market. The nature of these interactions are illustrated in Fig.~\ref{fig:Diagram}.

\subsection{Market Structure}
We consider a locational marginal price (LMP)-based wholesale electricity market that balances supply and demand at the minimum cost over multiple time periods subject to transmission constraints.  The set of time periods is denoted by $\mathcal{T}$. At a particular time $t$, the set of nodal demands is $\bd q_t$. We assume that the supply at each node $b$ and time period $t$ is characterized by an increasing function $c_{b,t}^\mathrm{gen}(\cdot)$ and is known by all participants. The system operator (SO) dispatches generation to minimize cost and the market clears at prices $\bd \lambda_t$.  Then, a participant that supplies $x$ amount of energy during period $t$ at node $b$ is paid $x \cdot \lambda_{b,t}$ and a participant consuming $y$ amount of energy is charged $y\cdot \lambda_{b,t}$.

\subsection{Energy Storage}

SUs perform energy arbitrage by shifting energy in time to take advantage of price differences~\cite{byrne2012estimating, Bradbury2014}.  Let $\mathcal{I}$ denote the set of $n$ SUs.  Let $d_{i,t}^+$ ($d_{i,t}^-$) denote the energy discharged (charged) by SU $i$ at time $t$.  We define $d_{i,t}=d_{i,t}^+ - d_{i,t}^-$.  A positive $d_{i,t}$ implies power injection while a negative $d_{i,t}$ implies power absorption.

Since the SUs interact with the market through an aggregator, the price they face could be different than the market prices $\bd \lambda_t$~\cite{KoponenEtAl2012}.  Let $\tau_{i,t}$ denote the price (\emph{e.g.}, retail rate) that SU $i$ faces at time $t$. Then, SU $i$'s profit is:
\begin{equation} \label{eqn:pi_s}
\pi_i^\mathrm{s}(\boldsymbol{d}_i ; \boldsymbol{\tau}_i) = \sum_{t \in \mathcal{T}} \tau_{i,t} \cdot d_{i,t}  - c_i(\boldsymbol{d}_i),
\end{equation}
where $\boldsymbol{d}_i  =\{d_{i,t}^+, d_{i,t}^- \}_{t\in\mathcal{T}}$ and $\btau_i=\left\{\tau_{i,t} \right\}_{t\in\mathcal{T}}$ denote a vector of storage actions and a vector of prices faced by SU $i$ throughout time, respectively.  The cost function $c_i(\cdot)$ models the degradation cost (due to cycling) of storage~\cite{Ortega-Vazquez}.  For chemistry-based energy storage, degradation is a function of internal chemical phenomena and dependent on several variables including the charge/discharge rate and depth of discharge~\cite{Koller}.  In this work, $c_i(\cdot)$ is a non-negative and strictly convex function of $\bd{d}_i$.  Strict convexity captures the fact that larger energy charges/discharges lead to higher degradation~\cite{ Ortega-Vazquez,Koller}.  Different cost models can be adopted within the proposed framework.

In this work, we consider additional revenue streams, (\emph{e.g.}, payments for frequency regulation and reserves~\cite{Xu_PESGM2016}), or benefits delivered by SUs (\emph{e.g.}, improvement of power quality~\cite{Saxena}) to be high-priority tasks modeled by constraints on the SU operation. For instance, a SU a can reserve some of its capacity for frequency regulation and contract the remainder for market participation through an aggregator.

The actions of the SUs are limited by a set of operating constraints. The most notable constraint of energy storage is that it must be \emph{energy neutral}, that is, the total energy charged over $\mathcal{T}$ is equal to the total energy discharged plus losses. The losses are modeled by discharge efficiency $\eta_i^+$ and charge efficiency $\eta_i^- $ coefficients.  The energy neutrality constraint models the fact that the SU participates in the market to perform energy arbitrage and does not generate net energy. The functions $\bd{h}_i (\bd{d}_i)$ denote the power, state-of-charge (SoC), and other constraints of SU $i$.

The profit maximization problem SU $i$ is expressed as
\begin{subequations} \label{eqn:opt}
	\begin{align}
	\max_{\bd d_i} \; &  \pi_i^\mathrm{s} (\bd{d}_i;\boldsymbol{\tau}_i ) \\
	\mbox{s.t. }  &\sum_{t \in \mathcal{T}}\left(\eta_i^- \cdot d_{i,t}^{-} - \frac{d_{i,t}^{+}}{\eta_i^+}\right)\! =\!0 \label{eq:energybalance}\\
	&\boldsymbol{h}_i(\boldsymbol{d}_i) \le 0  \label{eq:otherconst}\\
	& d_{i,t}^+ \geq 0 ,\; d_{i,t}^- \geq 0  \quad \forall t \in \T \label{eq:positivity}\\
	& \boldsymbol{d}_i  =\{d_{i,t}^+, d_{i,t}^- \}_{t\in\mathcal{T}},\nonumber
	\end{align}
\end{subequations}
where the objective is defined by Equation \eqref{eqn:pi_s}, \eqref{eq:energybalance} is the energy neutrality constraint; \eqref{eq:otherconst} includes the charge, discharge, SoC, and operating constraints; and \eqref{eq:positivity} are the non-negativity constraints.
		\subsection*{Constraints in the $\bd h_i(\cdot)$ function} The operating constraints embedded in the function $\bd h_i(\cdot)$ are the following:
		\begin{itemize}
			\item \emph{State-of-charge dynamics and limits}: The SoC of unit $i$ at time $t$ is defined as the amount of energy stored in the unit:
		\end{itemize}
		\begin{align*}
		s_{i,t} = s_{i,0} + \sum_{t' = 0}^{t-1}\left(\eta_i^- \cdot d_{i,t'}^{-} - \frac{d_{i,t'}^{+}}{\eta_i^+}\right).
		\end{align*}
		\begin{itemize}[label={}]
			\item Naturally, the SoC of a SU is constrained to be within a lower and an upper bound at all times:
		\end{itemize}
		\begin{align*}
		\underline{s}_i \le s_{i,t} \le \overline{s}_i \;  \forall\; t \in \T.
		\end{align*}
		\begin{itemize}
			\item \emph{Charge/discharge limits}: In addition to SoC constraints, the rate of charge and discharge of the SU is limited by upper and lower bounds. As noted by Equations~\eqref{eq:positivity}, the rate charge and discharge are lower bounded by $0$. The upper charge and discharge limits of SU $i$ are expressed as
		\end{itemize}
		\begin{align*}
		&d_{i,t}^- \le \overline{d}_{i}^-  \; \forall \; t\in\T \\
		&d_{i,t}^+ \le \overline{d}_{i}^+ \; \forall \; t\in\T.
		\end{align*}
		\begin{itemize}
			\item \emph{Other operating constraints}: Typical operating constraints of SUs employed in the literature are: the SoC limits, charge/discharge limits, and the energy neutrality constraint\footnote{The energy neutrality constraint is often expressed as a constraint restricting the final SoC to equal the initial SoC.}. In reality, however, SUs may operate under a larger set of operating constraints. For instance: chemistry-based batteries may be constrained by internal electrochemical processes~\cite{Lee2017}; the operation of the SU may be constrained by higher priority applications such as storing energy exclusively for emergency situations~\cite{Markkula} or improving power quality~\cite{Saxena}. The proposed SU model allows for arbitrary constraints to be embedded in the function $\bd h_i(\cdot)$.
		\end{itemize}

\subsection{Aggregator}
Most independent system operators require a minimum capacity from their participants.  For instance, ISO New England requires the participants of its wholesale market to have a capacity of at least 1 MW~\cite{ISONE_energy_storage}.  Therefore, small SUs such as EVs or household-sized energy storage can only participate in the market through aggregating entities. 

Let the vector $\bd d^\mathrm{bus}_t$ denote a vector of net nodal energy injections/absorptions by the SUs and be defined as  $\bd d^\mathrm{bus}_t=\sum_{i \in \mathcal{I}} \bd m^\mathrm{s}_i \cdot d_{i,t}$ where the $b^\mathrm{th}$ element of the vector $\bd m^\mathrm{s}_i$ is $1$ if SU $i$ is connected to bus $b$ and $0$ otherwise.  The $b^\mathrm{th}$ element of $\bd d^\mathrm{bus}_t$ denotes net discharge (if positive) or charge (if negative) of the SUs connected to node $b$. Then, the aggregator's net revenue from market participation is
\begin{equation*} 
\sum_{t \in T} { \bd \lambda_t^\top \bd d_t^\mathrm{bus}}.
\end{equation*}
In this work, we assume that an aggregator acts in a \emph{non-adversarial} manner, according to the following definition.
		\begin{defn}
			A player is said to be non-adversarial if it never chooses a strategy that results in a negative profit.
		\end{defn}
When $\bd d_t^\mathrm{bus} = \bd 0\; \forall\;t\in\mathcal{T}$, no energy is exchanged between market and aggregator; therefore the aggregator's net revenue is zero.  A non-adversarial aggregator implies that it does not choose actions if they result in a loss. Note that rational players\footnote{A rational player takes profit/utility maximizing decisions.} are also non-adversarial.

\begin{figure*}[h!]
	\centering
	\includegraphics[width=\textwidth]{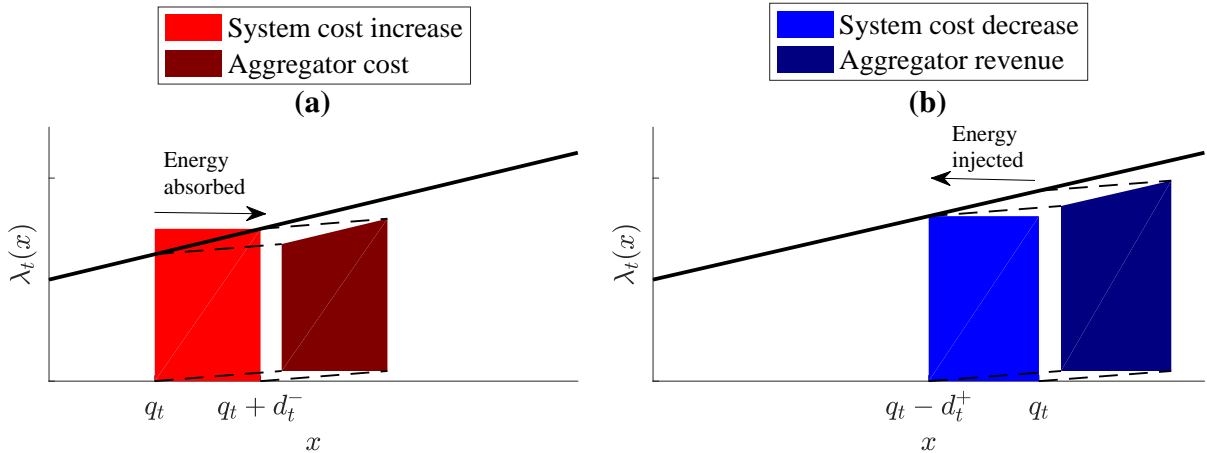}
	\caption{Subplot (a) shows a sample period when the net actions of the SUs result in energy absorption from the grid. Conversely, Subplot (b) shows a sample period when the net actions of the SUs result in energy injection to the grid.}
	\label{fig:illustration}
\end{figure*}
\section{Effect of the Aggregator on the System's Operating Cost}\label{sec:agg}

In this section, we show that the presence of a non-adversarial aggregator of SUs never increases the system cost. The system operating cost and LMPs are results of the following market clearing/system cost minimization problem.

		\begin{defn}
			Given a set of hourly energy storage actions $\bd d=\{\bd d_i\}_{ i \in \mathcal{I}}$ and hourly load $\bd q=\{ \bd q_t \}_{ t \in \T }$, the system operating cost is defined as:
			\begin{subequations} \label{eq:syst_prob}
				\begin{align}
				S(\bd q , \bd d) =& \min_{\bd g_t \ge 0} \sum_{b \in \mathcal{N}, t \in \T}\int_0^{g_{b,t}} \!\!\!c_{b,t}^\mathrm{gen}(x)dx+ \sum_{i \in \mathcal{I}} c_i(\bd d_i) \label{eqn:Sys_obj}  \\
				&\mbox{s.t.}  \nonumber \\
				& \bd 1^\top \bd g_t  -\bd 1^\top \bd q_t +\bd 1^\top \bd d_t^\mathrm{bus} =0 \;\;\; (\zeta_t)\; \forall \; t \in \T \label{eq:sys_power_balance}\\
				& \bd H(\bd g_t  - \bd q_t + \bd d_t^\mathrm{bus}) \le \overline{\bd f} \;\;\; (\bd \mu_t) \; \forall\; t \in \T \label{eq:sys_trans_const} \\
				&\bd d_t^\mathrm{bus} =\sum_{i \in \mathcal{I}} \bd m^\mathrm{s}_i \cdot (d_{i,t}^+ - d_{i,t}^-) \; \forall\; t \in \T \label{eq:d_bus_defn} \\
				& \boldsymbol{d}_i \! =\{d_{i,t}^+, d_{i,t}^- \}_{t\in\mathcal{T}} \;\forall\; i \in \mathcal{I}\label{eq:s_action_defn}
				\end{align}
			\end{subequations}
			where the objective~\eqref{eqn:Sys_obj} is to minimize the generation and energy storage costs during the horizon $\T$. Equations~\eqref{eq:sys_power_balance},~\eqref{eq:sys_trans_const}, represent the system power balance constraints and represent the system transmission constraints, respectively. Equations~\eqref{eq:d_bus_defn} define the net nodal injections/absorptions from the SUs and~\eqref{eq:s_action_defn} define the storage action vectors. The $b^\mathrm{th}$ entry of vector $\bd g_t$ represents generation from node $b$ at time $t$. The symbol $\bd H$ represents the transmission shift factor matrix and $\overline{\bd f}$ represents line limits. The symbols $\zeta_t$ and $\bd \mu_t$ are the dual variables of their respective constraints.
		\end{defn}
		\begin{defn} \label{defn:LMP}
			The LMPs at time $t$ are defined as the gradient of the Lagrangian of problem~\eqref{eq:syst_prob} with respect to the nodal demands at time $t$, i.e., $\bd \lambda_t = \nabla_{\bd q_t} \mathcal{L}$.
		\end{defn}

		This definition of the system cost and LMPs is standard in power system economics~\cite{KirschenEtAl2004c}.  The following theorem states that compared to a system without storage, a non-adversarial aggregator never increases the system cost regardless of how much market power an aggregator may have or exercise.

		\begin{thm} \label{thm:agg}
			Given a set of demands $\bd q$, and a vector of storage actions $\bd d$ from a non-adversarial aggregator, \vspace{-0pt}
			\begin{equation} \label{eqn:agg_welfare}
			S(\bd q,\bd d) \leq S(\bd q, \bd 0)
			\end{equation}
			where $S$ is the system operating cost defined by problem \eqref{eq:syst_prob}.
		\end{thm}

Theorem~\ref{thm:agg} implies that it is always beneficial to have SUs participating in the system, even when they are controlled by a price-anticipatory aggregator.  This in line with the expectations from the CAISO roadmap~\cite{CAISO2014}, and FERC Order 792~\cite{FERC2013} which introduces regulation to encourage the participation of small SUs.

Of course, if the aggregator has and exercises market power, the set of SU actions is socially suboptimal. In the case of generators, market power is typically exercised by overstating generation costs and/or understating available capacity.  In the case of energy storage performing arbitrage, market power can be exercised by \emph{withholding} storage services (\emph{i.e.}, shifting less than the socially optimal amount of energy). Section~\ref{sec:mitigation} presents a strategy to mitigate the adverse effects of market power via a non-uniform pricing scheme.
\subsubsection*{Informal proof of Theorem~\ref{thm:agg}}
Consider a one bus system. During any given time period, the SUs exhibit either net energy absorption or net energy injection. Fig.~\ref{fig:illustration}(a) shows a time period when the storage absorbs energy from the grid.  During this period, the system operating cost increases (as compared to a system with no storage) due to increased energy production used to supply the storage.  The system cost increase is illustrated by the red trapezoid. The aggregator's cost of purchasing energy is illustrated by the red rectangle.  Fig.~\ref{fig:illustration}(b) represents a time period when storage injects energy into the grid.  During this period the system operating cost decreases (as compared to a system with no storage) because the energy injections offset energy production. The cost decrease is illustrated by the blue trapezoid. The aggregator's revenue from selling energy is illustrated by the blue rectangle. When an aggregator arbitrages energy at a positive profit, 1) the blue rectangle is larger than the red rectangle and 2) the system cost blue trapezoid is larger than the red trapezoid. That is, when the aggregator operates at a profit, the system cost always decreases with respect to a system with no storage. The formal proof of Theorem~\ref{thm:agg} is available in the Appendix.

\section{Single-shot game}
\label{sec:seqgame}
Theorem~\ref{thm:agg} shows that having a non-adversarial storage aggregator is always beneficial to the system.  However, the actual actions of the SUs depend on the nature of the relationships between the aggregator and its SUs.  In this section, we present a Stackelberg game model for aggregator-SUs short-term interaction and show that the equilibria could be unsatisfactory for both parties.

\subsection{Aggregator-SUs Stackelberg Game}
The interaction between the aggregator and SUs is modeled as a single leader multi-follower game with perfect information~\cite{Cruz}.  The aggregator leads by announcing prices $\btau_i$ and the SUs respond by deciding their actions $\bd d_i$.

Let the feasible region of storage actions defined by Equations \eqref{eq:energybalance}--\eqref{eq:positivity} be denoted by $\mathcal{S}_i^\mathrm{s}$. For each price schedule $\btau_i$ that the aggregator chooses, each SU chooses a charge/discharge schedule
$\boldsymbol{d}_i^\mathrm{SE}=\argmax_{\boldsymbol{d}_i \in \mathcal{S}_i^\mathrm{s}} \pi_i^\mathrm{s}(\boldsymbol{d}_i ; \boldsymbol{\tau}_i) $ that maximizes its profit.  Since the SU profit maximization problem has a strictly concave objective, it yields a unique maximum $\boldsymbol{d}_i^\mathrm{SE}$.

The objective of the aggregator is to choose a set of prices $\btau=\left\{\btau_{i} \right\}_{i \in \mathcal{I}}$ that maximize its profit given by:
\begin{equation} \label{eqn:pi_a} \vspace{-0pt}
\pi^\mathrm{a}\!\left(\boldsymbol{\tau}; \boldsymbol{d}\right) \!= \sum_{i\in\mathcal{I}} \sum_{t\in\mathcal{T}} \left\{ \bd \lambda_t^\top \bd m^\mathrm{s}_i \cdot d_{i,t} -\tau_{i,t} \cdot d_{i,t}  \right\}.
\end{equation}
That is, it chooses  $\btau^\mathrm{SE} =\argmax_{0 \le \bd{\tau} \le M} \pi^\mathrm{a}(\boldsymbol{\tau} ; \{\boldsymbol{d}_i^\mathrm{SE}\}_{i \in \mathcal{I}})$.  For technical convenience, we assume that the aggregator can only set prices within a certain~(possibly large) bound $M$.  Then, there exists a (price , storage action) pair that constitutes the \emph{Stackelberg Equilibrium} for the aggregator-SUs game~\cite{Cruz}.
The Stackelberg Equilibrium, $(\btau^\mathrm{SE},\bd d^\mathrm{SE})$, can be obtained by formulating the aggregator's problem as a mathematical problem with equilibrium constraint (MPEC). In general, MPECs are computationally difficult non-linear problems.  Nonetheless, efficient algorithms to solve some MPECs are available.  See references~\cite{Momber_Wogrin_2016,Weywadt2012,Ferris} for further details.

A natural question about the Stackelberg Equilibrium is how \emph{efficient} it is.  Here, we measure efficiency via the notion of Pareto-optimality.
\begin{defn}
	A strategy $A$ is Pareto-superior to a strategy $B$ if at least one player is strictly better off with $A$ and no one is worst-off.
	We then say a strategy is Pareto-inefficient if there exists a Pareto superior strategy.
\end{defn}
The following example shows that the equilibrium could be Pareto-inefficient.

\begin{ex} \label{Ex:equential} In this, and throughout the rest of the examples, we assume that an aggregator interacts with one SU ($n=1$) in a single-node system and that the time horizon is composed by two time periods ($n_t=2$). The price at each time period is given by $\lambda_t(x_t)=x_t\;\forall \; t=1,2$ where $x_t$ is the net load at time $t$.  The net load is $-d_1$ when $t=1$ and $5 - d_2$ when $t=2$.  The round-trip efficiency of the SU is $\eta=0.95$, its charge and discharge limits are $1$, its SoC must be in $[0,1]$ at all times, and its cost function is given by $c(\boldsymbol{d}) = \frac{1}{2}\cdot\sum_{t=1}^2  d_{t}^2$. We consider a single SU and a single node to reduce notational clutter and suppress the indices $i$ and $b$.

	When the aggregator and SU play a single-shot game, their actions are as follows.  The aggregator sends a price schedule such that $\Delta \tau^\mathrm{\;SE}=\tau_1^\mathrm{\;SE}- \eta \cdot \tau_2^\mathrm{\;SE} = -1.19$ while the SU charges $0.62$ units of energy during the first time period.  Because of the energy neutrality requirement for the SU, it discharges $0.95\cdot 0.62$ units of energy during the second time period.  The Stackelberg profits are then ${\pi^\mathrm{a}}^\mathrm{\;SE}=1.48 $ and ${\pi^\mathrm{s}}^\mathrm{\;SE} =0.37$, respectively.  However, suppose that the aggregator and SU agree to the following:
	The SU charges $0.7$ units of energy during the first time period and the aggregator sets a price schedule such that $\Delta \tau=\tau_1- \eta \cdot \tau_2 = -1.25$.  This strategy is Pareto-superior to the Stackelberg Equilibrium:  both the SU's and the aggregator's profits are higher than the Stackelberg profits (see Fig. \ref{fig:ex1}).
\end{ex}

Any strategy other than a Stackelberg Equilibrium, however, is necessarily unstable.  Thus, cooperation during a single-shot game is not necessarily compatible with self-interested players. It is also important to note that the overall system cost may suffer from adversarial relationships between aggregators and SUs, since the actions from SUs may be artificially suppressed. However, the aggregator and SU may be in a position where they interact repeatedly over time.  In this case, they may be induced to cooperate and adopt equilibria that are at the very least Pareto-superior to the Stackelberg Equilibrium.

\section{Long-term cooperation}  \label{section:longtermcoop}
We showed that the Stackelberg Equilibrium is not necessarily Pareto-efficient and that other solutions are not stable.  In practice, this view is pessimistic: most aggregators and SUs do not play a single-shot game.  Instead, they interact repeatedly and their strategies during these games determine the \emph{long-term} outcome.  In such cases, an aggregator and SUs may reach a cooperative equilibrium that is stable and Pareto-superior with respect to the Stackelberg Equilibrium.  This section establishes conditions of stable cooperative equilibria.
\begin{rmk}
 It is trivial to show that all cooperative equilibria are Pareto-superior to the Stackelberg Equilibrium since a worst-off player could repeatedly play the Stackelberg Equilibrium.
\end{rmk}

\subsection{Repeated game model}
The long-term profit of each player is defined as the discounted sum of the single-shot profits.  Here, we assume that the exact same single-shot game is played repeatedly and that market prices of future games can be thought of as forecasted prices.  The long-term profit of the aggregator is denoted by \vspace{0pt}
\begin{align*} \vspace{0pt}
\pi^{\mathrm{a},\infty} = \sum_{k=0}^{\infty}{ \delta^k \cdot\pi^\mathrm{a}(\boldsymbol{\tau}^{(k)}; \boldsymbol{d}^{(k)})  }
\end{align*}
where $ \delta \in (0,1)$ is the discount rate\footnote{The discount rate reflects the time value of money.}.  The symbols $\boldsymbol{\tau}^{(\!k\!)}$ and $\boldsymbol{d}^{(\!k\!)}$ denote strategy decisions for the $k^\mathrm{th}$ time the single-shot game is played.
Similarly, the long-term profit of SU $i$ is denoted by
\begin{align*}\vspace{0pt}
\pi^{\mathrm{s},\infty}_i = \sum_{k=0}^{\infty}{ \delta^k \pi_i^\mathrm{s} (\boldsymbol{d}_i^{(k)} ; \boldsymbol{\tau}_i^{(k)} ) } .
\end{align*}

Next, we identify the strategy spaces of the aggregator and SUs. Specifically, we state the cooperation and defection strategies.
\subsubsection{Cooperation strategies}
The cooperation strategy of the aggregator,
\begin{align*}
\boldsymbol{\tau}_i^{(\!k\!)}=\begin{cases}
\widehat{\boldsymbol{\tau}}_i \;\; \mathrm{if}\;\; \boldsymbol{d}^{(m)}_i=\widehat{\boldsymbol{d}}_i \quad \forall \; m <k \\
\btau_i^\mathrm{D}\;\; \text{otherwise}
\end{cases} \;\forall\; i \in \mathcal{I},
\end{align*}
describes the strategy in which during the $k^\mathrm{th}$ game, the aggregator sends SU $i$ previously agreed prices $\widehat{\btau}_i$ if the SU has played agreed actions $\widehat{\bd{d}}_i$ during all previous times. If the SU failed to uphold its commitment during the previous game,  the aggregator plays the defection strategy $ \btau_i^\mathrm{D}$ (defined shortly). Likewise,
\begin{align*}
\boldsymbol{d}_i^{(\!k\!)}=\begin{cases}
\widehat{\boldsymbol{d}}_i \;\; \mathrm{if}\;\; \boldsymbol{\tau}_i^{(m)}=\widehat{\boldsymbol{\tau}}_i \quad \forall \; m \le k \\
\boldsymbol{d}_i^\mathrm{D} \;\; \text{otherwise}
\end{cases}
\end{align*}
describes the cooperation strategy of SU $i$. The SU plays agreed actions $\widehat{\boldsymbol{d}}_i$ if the aggregator has upheld its commitment to send agreed prices $\widehat{\boldsymbol{\tau}}_i$ during all previous times the game has been played. Otherwise, the SU $i$ plays the defection strategy $\boldsymbol{d}_i^\mathrm{D}$ for the subsequent times the game is played.

\subsubsection{Defection strategies}
\label{subsubsection: Defection S}

The SU defects by maximizing its single-shot profit.  Its defection strategy is given by $\boldsymbol{d}_i^\mathrm{D} =\argmax_{\boldsymbol{d}_i \in \mathcal{S}_i^\mathrm{s}} \pi_i^\mathrm{s}(\boldsymbol{d}_i ; \widehat{\btau}_i)$.  Similarly, the aggregator maximizes its profit derived from SU $i$ by playing $\boldsymbol{\tau}_i^\mathrm{D}=\argmax_{0 \le \boldsymbol{\tau}_i \le M} \pi^\mathrm{a}(\{\boldsymbol{\tau}_i,\widehat{\btau}_{-i}\} ; \widehat{\bd{d}})$ where the subscript $-i$ represents all SUs except $i$.

\subsubsection{Defection equilibrium}
\label{subsubsection: Defection eq}
From the definition of the cooperation strategies, once the aggregator defects from cooperation with SU $i$, the SU also defects.  Similarly, if SU $i$ defects, the aggregator defects form cooperation with SU $i$ in the following single-shot game.  Thus, they fall into playing a single leader, single follower sequential game while the actions of the rest of the SUs remain fixed.  The Stackelberg Equilibrium of this single leader, single follower game is given by $\boldsymbol{d}_i'=\argmax_{\boldsymbol{d}_i \in \mathcal{S}_i^\mathrm{s}} \pi_i^\mathrm{s}(\boldsymbol{d}_i ; \btau_i)$ and $\boldsymbol{\tau}_i'=\argmax_{ 0 \le \boldsymbol{\tau}_i \le M} \pi^\mathrm{a}(\{\boldsymbol{\tau}_i,\bd{\tau}_{-i}\} ; \{\bd{d}_i',\bd{d}_{-i}\})$.  Denote the defection equilibrium profit of the aggregator as ${\pi_i^\mathrm{a}}'$ and the defection equilibrium profit of the SU as ${\pi_i^\mathrm{s}}'$.


\subsection{Ensuring cooperation in an infinitely repeated game}
\label{subsec:Ensuringcoop}
This subsection characterizes the set of strategies that ensures cooperation by rational players in the long-term.  Then we use the Nash Bargaining Model to conclude that the profit split is Pareto-efficient (\emph{i.e.}, the aggregator and SUs cooperate to maximize their aggregate profit) and to predict precisely how the aggregate profit is shared among all players.

Given a set of agreed prices $\widehat{\btau}_i$ and storage actions $\widehat{\bd{d}}_i$, all players decide whether (and when) to defect by maximizing their long-term profits. A player cooperates if no finite time of defection maximizes its long-term profit. A set of agreed prices and storage actions is a cooperative equilibrium in the infinitely repeated game if they incentivize both the aggregator and SUs to never defect.
\begin{lem} \label{lem:coop}
 The aggregator cooperates with SU $i$ if its agreed profit is greater than its defection equilibrium profit ${\pi_{i}^\mathrm{a}}'$.  SU $i$ cooperates with the aggregator when its agreed profit is greater than \begin{align*}(1\!-\!\delta)\pi_{i}^\mathrm{s}(\bd{d}_i^\mathrm{D}; \btau_i)\!+\! \delta {\pi_{i}^\mathrm{s}}'.\end{align*}  The set of agreed profits that incentivize both the SU and the aggregator to cooperate define the cooperative equilibria.
\end{lem}

The proof of Lemma~\ref{lem:coop} is included in the Appendix.  As shown in the proof, infinitely many profits splitting schemes foster long-term cooperation by both parties.  As shown shortly, Nash Bargaining Theory is used to predict \emph{i)} precisely how the profits will be split and \emph{ii)} that the negotiation region is the set of profits that foster long-term cooperation.

 \begin{figure}[tb]
 \centering
 \includegraphics[width=.5\textwidth]{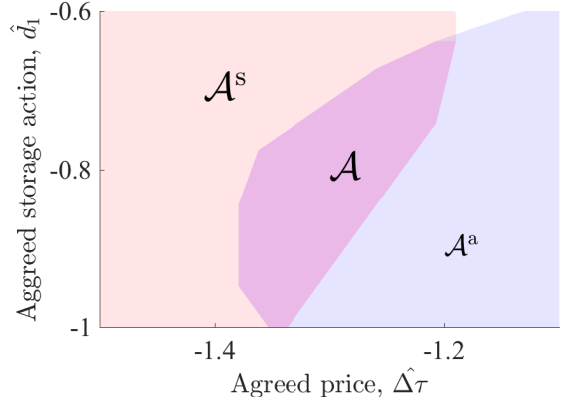}
 \caption{The x-axis shows the price set by the aggregator and the y-axis shows the storage actions~(charging amount).  If the actions are in the region $\mathcal{A}^\mathrm{s}$, the SU would never defect. If the actions are in $\mathcal{A}^\mathrm{a}$, the aggregator would never defect.  Then if the actions are in $\mathcal{A}=\mathcal{A}^\mathrm{s} \cap \mathcal{A}^\mathrm{a}$, both players are incentivized to cooperate indefinitely.} \vspace{-0pt}\label{fig:coop_reg}
\end{figure}

\begin{ex} This continues the previous example.  Let the discount rate be $\delta = 0.98$ and let
 \begin{align*} \vspace{0pt}
 \alpha_\mathrm{a}(\widehat{\Delta \tau} ,\widehat{d}_{1} )\!=   \pi^\mathrm{a}\!\!\left(\!\widehat{\boldsymbol{\tau}}; \widehat{\boldsymbol{d}}\!\right)-{\pi^\mathrm{a}}'.
 \end{align*}
 From Lemma \ref{lem:coop}, the aggregator is incentivized to cooperate during every single-shot game when $\alpha_\mathrm{a}(\widehat{\Delta \tau} ,\widehat{d}_{1} ) \ge 0$. Similarly, let
 \begin{align*}
 \alpha_\mathrm{s}(\widehat{d}_{1},\widehat{\Delta \tau}  ) =   \pi^\mathrm{s}\!\left(\!\widehat{\boldsymbol{d}}; \widehat{\boldsymbol{\tau}}\!\right)-\!(1\!-\!\delta)    \pi_{i}^\mathrm{s}(\bd{d}_i^\mathrm{D}; \widehat{\btau}_i)  \!-\!\delta{\pi^\mathrm{s}}' \end{align*}
 Also from Lemma \ref{lem:coop}, the SU is incentivized to cooperate when $\alpha_\mathrm{s}(\widehat{d}_{1},\widehat{\Delta \tau}  ) \ge 0$.

 Let the sets $\mathcal{A}^\mathrm{s} = \{ ( \widehat{d}_{1},\widehat{\Delta \tau} )  | \alpha_\mathrm{s}(\widehat{d}_{1},\widehat{\Delta \tau}  ) \ge 0\}$, $\mathcal{A}^\mathrm{a}= \{ ( \widehat{d}_{1},\widehat{\Delta \tau} )  | \alpha_\mathrm{a}(\widehat{\Delta \tau}, \widehat{d}_{1} ) \ge 0\}$ denote the SU cooperation region and the aggregator cooperation region, respectively.  Set $\mathcal{A}^\mathrm{s}$ represents all the charging/price pairs that incentivize the SU to never fall back to the Stackelberg Equilibrium and uphold its cooperation agreement with the aggregator. The same is true for the set $\mathcal{A}^\mathrm{a}$ with respect to the aggregator. The region where both players cooperate is denoted by $\mathcal{A} =\mathcal{A}^\mathrm{s} \cap \mathcal{A}^\mathrm{a}$.  The sets $\mathcal{A}^\mathrm{s}$, $\mathcal{A}^\mathrm{a}$, and $\mathcal{A}$ are illustrated in Fig. \ref{fig:coop_reg}.

\end{ex}

\subsection{Profit split via Nash Bargaining}
\label{section:NB}
Lemma~\ref{lem:coop} shows that there is a range of strategies that ensure long-term cooperation,\emph{i.e.}, there are many cooperative equilibria.  In practice, one of these equilibria needs to be chosen.  Knowing precisely which equilibrium strategy is to be played is crucial to predicting the aggregator's market participation strategy (\emph{e.g.}, its supply/demand bids).  This section presents the Nash Bargaining Model~\cite{Nash-1950} used to predict said long-term equilibrium.


Let  $\mathcal{B}_i = \mathcal{S}^\mathrm{s}_i\times \mathcal{S}^\mathrm{a}_i$ denote the set of possible outcomes of a bargaining process between the aggregator and SU $i$ and the outcome as $\left(\widehat{\boldsymbol{d}}_i^*, \widehat{\boldsymbol{\tau}}_i^* \right)=\xi\left(\mathcal{B}_i\right)$.  The function $\xi(\cdot)$ maps a set of possible outcomes to the solution to the bargaining problem~\cite{Nash-1950}. Here, $\mathcal{S}^\mathrm{a}_i = \{ \btau_i | 0 \le \btau_i \le M\}$.  The Nash Bargaining Solution is a single point in a set of possible outcomes that satisfies the following axioms:
\begin{itemize}
 \item \emph{Pareto-efficiency}: If $\boldsymbol{a},\;\boldsymbol{b} \in \mathcal{B}_i$, $\pi^{\mathrm{a}}(\boldsymbol{a})>\pi^{\mathrm{a}}(\boldsymbol{b})$, and $\pi^{\mathrm{s}}_i(\boldsymbol{a})>\pi^{\mathrm{s}}_i(\boldsymbol{b})$ then $\boldsymbol{b} \ne\xi\left(\mathcal{B}_i\right)$.
 \item \emph{Independence of irrelevant alternatives}: If $\tilde{\mathcal{B}}_i \subseteq \mathcal{B}_i$ and $\xi\left(\mathcal{B}_i\right) \in \tilde{\mathcal{B}_i}$, then $\xi(\tilde{\mathcal{B}}_i) = \xi\left(\mathcal{B}_i\right)$.
\end{itemize}
\begin{lem}\label{lem:bargain}
 Assume that the aggregator engages in bilateral negotiations with each SU and that all players are risk neutral\footnote{As shown by~\cite{Binmore}, if one of the players is more risk adverse than the other, its share of the profit decreases.  Conversely, if a player is more risk-loving than the other, its share of the profit increases.}.  The aggregator and SUs split the maximum aggregate profit given by:
 \begin{align}
 \pi(\boldsymbol{d}^*) = \max_{\bd{d} \in \mathcal{S}^\mathrm{s}} \{ \pi^\mathrm{a}(\btau; \bd{d}) +\sum_{i\in \mathcal{I}} \pi_i^\mathrm{s}(\bd{d}_i ; \btau_i) \}. \label{eq:Marketclear}
 \end{align}
 where $\mathcal{S}^\mathrm{s}=\mathcal{S}^\mathrm{s}_1 \times\hdots \times \mathcal{S}^\mathrm{s}_n$. The unique Nash Bargaining Solution is given by:
 \begin{align}
 \forall i:(\widehat{\bd d}^*_i, \widehat{\btau}^*_i)= \argmax_{\substack{ 0 \le \boldsymbol{\tau}_i \le M \\ \bd d_i = \bd d_i^* \\ (\boldsymbol{\tau}_i, \boldsymbol{d}_i) \in \mathcal{A}_i }} &{ \left(\pi_i^\mathrm{s} - {\pi_i^\mathrm{s}}'  \right)\left(\pi^\mathrm{a} - {\pi_i^\mathrm{a}}'  \right) } \label{eq:bargprob}
 \end{align}
 where $\pi^\mathrm{a} \equiv \pi^\mathrm{a} (\{ \boldsymbol{\tau}_i,\widehat{\btau}_{i-}^* \}; \{ \boldsymbol{d}_i,\widehat{\bd d}_{i-}^* \})$ and $\pi_i^\mathrm{s} \equiv \pi_i^\mathrm{s} (\boldsymbol{d}_i; \boldsymbol{\tau}_i)$.  The set of agreed prices and storage actions that deliver profits in the cooperative equilibria is denoted by  $\mathcal{A}_i$.
\end{lem}

The proof of Lemma~\ref{lem:bargain} is available in the Appendix.  This result is important because it allows us to simplify the aggregator-SU model and treat the aggregator as an entity who bids the solution of problem~\eqref{eq:Marketclear}, $\bd d^*$, into the market.  Moreover, this model allows us to study how the aggregator bid $\bd d^*$ compares to the socially optimal behavior of the SUs.

The next section compares the system cost when the aggregator bids $\bd d^*$ with the socially optimal system cost.  Since the aggregator is price-anticipatory, it is able to exercise market power by deviating from socially optimal bids in order to increase its profit.  Then, we propose a non-uniform pricing scheme that incentivizes the aggregator to behave in a socially optimal manner.

\begin{ex}  \label{ex:bargaining}
 This continues the previous example.  Suppose that the aggregator engages in negotiations with the SU.  By the Pareto-efficiency axiom, the aggregator and SU agree on storage actions $\widehat{d_{1}}^*=d_{1}^*=-0.83$ which yield the maximum aggregate profit.  Also from the Pareto-efficiency axiom, the agreed price schedule $\widehat{\Delta \tau}^*$ will be one that fosters long-term cooperation.  From the previous example, long-term cooperation is fostered when
 \begin{align*}
 \alpha_\mathrm{a}(\widehat{\Delta \tau}, -0.83  ) \ge 0 \; \mathrm{and} \;\alpha_\mathrm{s}(-0.83,\widehat{\Delta \tau}  ) \ge 0
 \end{align*}
 or equivalently, when $ -1.23 \le \widehat{\Delta \tau} \le -1.37$.

 From reference \cite{Binmore}, the bargaining outcome is given by the solution to the quadratic optimization problem: \vspace{0pt}
 \begin{align*} \vspace{0pt}
 \max_{-1.23 \le \widehat{\Delta \tau} \le -1.37} \!\! \left(\pi^\mathrm{s} (-0.83; \widehat{\Delta \tau}) \!-\! {\pi^\mathrm{s}}'\right) \left( \pi^\mathrm{a} ( \widehat{\Delta \tau}; \!-0.83)  \!-\! {\pi^\mathrm{a}}'\right)
 \end{align*}
 which is $\widehat{\Delta \tau}^*= -1.31$.

 Since $(\widehat{d_1}^*,\; \widehat{\Delta \tau}^*)$ lies on the interior of $\mathcal{A}$, the solution to the Nash Bargaining Problem is given by the solution to the system of equations $\pi^{\mathrm{s}} =-\pi^{\mathrm{a}} + \pi(d_1^*)$ and $\pi^{\mathrm{s}}=\pi^{\mathrm{a}}+{\pi^\mathrm{s}}' - {\pi^\mathrm{a}}'$ as shown in \cite{Nash-1950}.  This concludes the example.

 \begin{figure}[tb]
   \centering
   \includegraphics[width=0.5\textwidth]{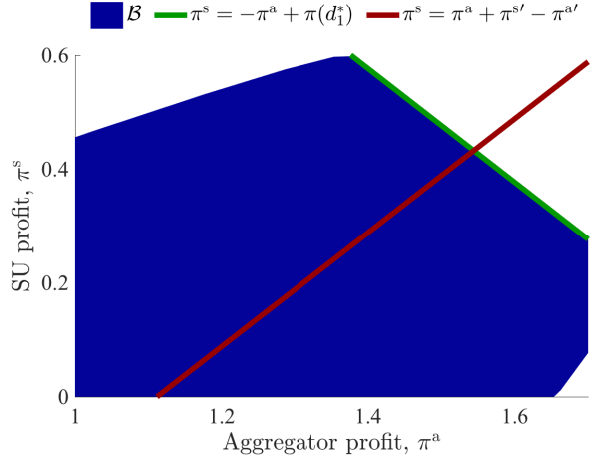}
   \caption{The set of possible outcomes bargaining outcomes is denoted by the area $\mathcal{B}$.  Each point inside this area represents a realizable profit split between the aggregator and SU.  The green line represents all possible ways of splitting the Pareto optimal (maximum) profit.  The portion of $\mathcal{B}$ that overlaps the green line represents the Pareto front.  The red line represents the line of symmetry.  The intersection of these two lines is the bargaining solution.} \label{Figure4}
 \end{figure}

\end{ex}

Note that for larger aggregator defection equilibrium profits ${\pi^\mathrm{a}}'$, the bargaining solution delivers a larger aggregator profit $\widehat{\pi}^{\mathrm{a}*}$ (\emph{i.e.}, the lines in Fig.~\ref{Figure4} intercept at a higher $\widehat{\pi}^{\mathrm{a}*}$).  Since ${\pi^\mathrm{a}}'$ is the ``outside option" profit of the aggregator (\emph{i.e.}, profit of the aggregator in case the negotiations fall apart), a larger outside option profit can be interpreted as the aggregator having greater bargaining power or ``leverage."  The same can be said for the relationship between the SU profit $\widehat{\pi}^{\mathrm{s}*}$ and the SU defection equilibrium profit ${\pi^\mathrm{s}}'$.

\section{Market power mitigation}
\label{sec:mitigation}
The Nash Bargaining Solution~\eqref{eq:bargprob} predicts that the aggregator bids $\bd{d}^*$. This allows us to abstract the aggregator-SU interactions and model the aggregator as an entity that bids  $\bd{d}^*$. However, this bid might not be socially optimal as the aggregator behaves in a price-anticipatory manner and might control enough capacity to have some degree of market power. This section proposes a non-uniform pricing mechanism designed to mitigate the market power of the aggregator.
\subsection{Social optimum}
A system operator (SO) would operate the SUs such that the system cost is minimized by solving:
\begin{align}
&\boldsymbol{d}^\mathrm{social}=\argmin_{\boldsymbol{d} \in \mathcal{S}^\mathrm{s} } {   S(\bd{q} , \bd d)}\label{eq:dictatorprob}.
\end{align}
The solution $\boldsymbol{d}^\mathrm{social}$ denotes the socially optimal storage actions.

\begin{ex} This continues the previous example. Note from Fig. \ref{profit_and_SW} that the maximum aggregator-SU profit peaks at an absolute value of $d_1$ that is lower than the minimum system cost. This suggests that the aggregator may want to withhold storage services to maximize profits.

The total system cost, given by
     \begin{align*}
\int_0^{-d_1} \!\!\!x\;dx +\int_0^{5-d_2}\!\!\! x\;dx   + c(\bd d),
\end{align*}
is minimized  when $d_1=-1$ and $d_2 = 0.95$. In this case, the total system cost is $9.65$ while the aggregator and SU split a profit of $1.89$. In the socially optimal case, the load pays $20.3$ for their consumption.  On the other hand, the market clears and the aggregator-SU maximizes its profit when $d_1^* =-0.83$ (point B in Fig. \ref{profit_and_SW}). In this case, both the system cost and the aggregator profits are higher. The exercise of market power by the aggregator hurts the load significantly as shown in Table \ref{table1}. This result motivates market power-mitigating mechanism introduced in the following subsection. This concludes the example.

\begin{table}[tb]
    \renewcommand{\arraystretch}{1}
    \caption{}
    \label{table1}
    \centering
    \bgroup
    \def\arraystretch{1}
    \begin{tabular}{|c| |c| c| c| }
        \hline
        Storage level & SU/agg. profit & Sys. cost & Load payment \\
        \hline
        Social optimum ($d_1^\mathrm{social}$) & $1.89$ & $9.65$ & $20.3$ \\
        Market clearing ($d_1^*$) & $1.98$ & $9.86$ & $21.1$ \\
        \hline
        \multicolumn{4}{l}{$d_1^\mathrm{social} = -1$, $d_1^* = -0.83$} \\
    \end{tabular}
    \egroup
\end{table}

\end{ex}


\vspace{0pt}
\subsection{Market clearing with market power mitigating price}
In order to operate the system at minimum cost, the SO could charge the aggregator a ``market power mitigating price'' (MPMP). The MPMP incentivizes a profit maximizing aggregator to behave in a socially optimal manner.
    \begin{thm} \label{thm:mpmp}
        If the aggregator pays the MPMP
        \begin{align*} \vspace{0pt}
        p_{b,t}(d_{b,t}^\mathrm{bus})
        \!= \!\frac{-1}{d_{b,t}^\mathrm{bus} } \!\cdot\!\left(\int_0^{g_{b,t}} \!\!\!\!c_{t,b}^\mathrm{gen}(x) dx -C_{b,t}\right)\; \forall \; b \in \mathcal{N},\; t \in \mathcal{T}
        \end{align*}
       in lieu of the market clearing price, the market clears at the social optimum. The constants $C_{b,t}$ can be used to regulate the aggregator's profit.
    \end{thm}
The proof of Theorem \ref{thm:mpmp} can be found in the Appendix. The proposed MPMP has three key properties. First, it is compatible with both rational behavior of the aggregator and the desire to maximize social welfare: by acting in a selfish, profit maximizing way, the aggregator carries out socially desirable actions. Second, the term $C_{b,t}$ allows for regulation of the aggregator's profit. And third, its formulation requires no private information: the characteristics (\emph{i.e.}, operating constraints and cost functions) of the SUs do not need to be disclosed.

\begin{ex}
    This continues the previous example. From Example~\ref{Ex:equential}, under traditional LMPs the aggregator is exposed market clearing prices $\lambda_1(-d_1) = -d_1$ and $\lambda_2(5-d_2) = 5 -d_2$. Under the MPMP, the aggregator is exposed to prices
    \begin{align*}
    &p_1(d_1) \!= -\frac{1}{d_1} \left(\int_0^{-d_1} \!\!\!\!\!\!x\; dx - C_1 \right)\!  =  -\frac{d_1^2 - 2C_1}{2d_1}\; and \\
    &p_2(d_2) \!= -\frac{1}{d_2} \left(\int_0^{5-d_2}  \!\!\!\!\!\!x\; dx - C_2\right)\! =  -\frac{(d_2-5)^2 - 2C_1}{2d_2}.
    \end{align*}
    Let $C= C_1 + C_2$. Then, when the aggregator is exposed to the MPMP, it maximizes its profit given by
    \begin{align}
    \sum_{t=1}^2 \!d_t \cdot p_t(d_t)   - c(\bd d) \!= -\frac{( d_2 - 5) ^2 + d_1^2}{2}-c(\bd d)+C. \label{eq:MPMP_obj}
    \end{align}
     Note that maximizing Equation~\eqref{eq:MPMP_obj} is equivalent to minimizing the system cost given by
     \begin{align}
     \int_0^{-d_1} \!\!\!x\;dx +\int_0^{5-d_2}\!\!\! x\;dx   + c(\bd d)  =  \frac{( d_2 - 5) ^2 + d_1^2}{2}+c(\bd d).\label{eq:SW_obj}
     \end{align}
     Even though Equations~\eqref{eq:MPMP_obj} and~\eqref{eq:SW_obj} are opposite in signs, they are an equivalent objectives since~\eqref{eq:MPMP_obj} is maximized and~\eqref{eq:SW_obj} is minimized. Then, when exposed to the MPMP, the objective of a profit-seeking aggregator (maximize profit) and the system operator (minimize cost) are exactly aligned. Thus, the aggregator bids the social optimum. This concludes the example.
\end{ex}

\subsection{Regulated profit of an aggregator}  \label{subsec:regulated_profit}
The proposed MPMP allows for the regulation of the aggregator's profit by selecting constants $C_{b,t}$ such that the aggregator receives a desired (and arbitrary) amount of profit. Such profit must come from charges to other players in the system. We assume that the profit allocation decision (\emph{e.g.}, deciding the profit level of the aggregator and the sources of such profits) is a problem to be solved by the regulator, the SO, market operator, and/or other stakeholders.  This, however, is a decision that is completely system, market structure, and regulatory environment-dependent.

 The amount of regulated profit does not affect the short-term operation of the SUs as the $C_{b,t}$'s come in as constants in the objective of problem \eqref{eq:Marketclear}.  However, selecting a proper amount of regulated profit is crucial as it may:
 \begin{enumerate}
 \item (over/under)-incentivize SUs to (come into/leave) the market if the regulated profit is too (high/low),
 \item (over/under)-incentivize other consumers or producers to (come into/leave) the market if the aggregator regulated profit is too (low/high).
\end{enumerate}

Another important implication of the regulated profit is how it will impact the profit splitting scheme given by the Nash Bargaining Solution.  The essence of the Nash Bargaining Solution is as follows: the aggregator and SUs will split profits in a symmetric manner. Everything equal (\emph{e.g.}, bargaining leverage and risk profiles), the aggregator will retain one half of the aggregate profit and distribute the remaining half among the SUs. Thus, a higher regulated profit higher profit for the aggregator translates into higher profits for the SUs. Conversely, a lower regulated profit decreases the ``size of the pie" to be shared among SUs and aggregator. The next numerical example illustrates the relationship between the regulated profit and the Nash Bargaining Solution.

\begin{ex}
    This continues the previous example.  As shown in Example~\ref{ex:bargaining}, the Nash Bargaining Solution is given by the following system of equations:
    \begin{subequations} \label{eq:nash_barg_sol}
    \begin{align}
&    \pi^{\mathrm{s}} =-\pi^{\mathrm{a}} + \pi(d_1^*) \\
&  \pi^{\mathrm{s}}=\pi^{\mathrm{a}}+{\pi^\mathrm{s}}' - {\pi^\mathrm{a}}'.
    \end{align}
        \end{subequations}
    Recall that $\pi(d_1^*)$ represents joint aggregator/SU profit when optimally bidding in a traditional LMP market.  Under the MPMP, however, the optimal bid is the social optimum and the profit is set to an arbitrary level via the $C$ constant.

    Suppose the regulator chooses a constant $C$ such that the joint aggregator-SU profit becomes a desired profit $\pi^\mathrm{reg}(C)$. Replacing $\pi(d_1^*)$ with $\pi^\mathrm{reg}(C)$; and replacing ${\pi^\mathrm{a}}'$ and ${\pi^\mathrm{s}}'$ with their respective numerical values, Equations~\eqref{eq:nash_barg_sol} become
        \begin{align*}
        &    \pi^{\mathrm{s}} =-\pi^{\mathrm{a}} + \pi^\mathrm{reg}(C)\\
        &  \pi^{\mathrm{s}}=\pi^{\mathrm{a}} -1.11 .
        \end{align*}
        Then, the profits of the SU and aggregator as a function of the regulated profit are
                \begin{align*}
            \pi^{\mathrm{s}} =\frac{ \pi^\mathrm{reg}(C)-1.11}{2} \mbox{ and } \pi^{\mathrm{a}} =\frac{ \pi^\mathrm{reg}(C)+1.11}{2},
        \end{align*}
        respectively.
        It is easy to see that if the regulator increases the regulated joint aggregator-SU profit by one unit, the aggregator and SU will split that unit equally, i.e.,
        \begin{align*}
        \frac{\partial     \pi^{\mathrm{a}}}{\partial     \pi^{\mathrm{reg}}} =\frac{\partial     \pi^{\mathrm{s}}}{\partial     \pi^{\mathrm{reg}}} = \frac{1}{2}.
        \end{align*}
        This concludes the example.
\end{ex}

\subsection{How does the MPMP fit in current market designs?}  \label{subsection:MPMP_current_mkt}
The MPMP has three major features that relate to the design of modern electricity markets. First, it encourages a monopolist aggregator of energy storage to bid the social optimum. Second, through its $C_{b,t}$ constants, it allows the regulator to assign the aggregator an arbitrary amount of profit. Last, as consequence of the profit regulation property, the MPMP could serve as a welfare allocation tool to accomplish regulatory goals, subsidize, or tax certain technologies in the power system. These three properties are compatible with current market designs and practices as discussed in the rest of this section.
\subsubsection{Market power mitigation}
One of the responsibilities of the Federal Energy Regulatory Commission (FERC) in the United States is to oversee wholesale electricity markets. FERC order 888~\cite{FERC888} states that
\begin{quote}
\emph{``The Commission's goal is to remove impediments to competition in the wholesale bulk power marketplace and to bring more efficient, lower cost power to the Nation's electricity consumers."}
\end{quote}
Aligned to FERC order 888, Independent System Operators throughout the country (\emph{e.g.}, CAISO, NE-ISO) have established departments dedicated to monitoring market activity and countering abuses of market power. Understandably, in both academia and in practice, such efforts have typically been focused on the market power of generators. Nevertheless, the proposed MPMP is in line with the goal of fomenting market efficiency and discouraging monopolistic behavior.
\subsubsection{Profit regulation}
The MPMP also has the ability to regulate the profit of a monopolist aggregator. Regulating profits of monopolies is a fairly common practice in many sectors of the economy, including the electricity sector. In fact, one of the most prominent examples of this is the profit regulation of electric and gas utilities~\cite{RAP}. A common way of regulating the profit of electric utilities is through \emph{cost-plus} schemes where the utilities are allowed to cover their costs plus an administratively designated rate of return.
\subsubsection{Welfare allocation}
The MPMP can also play alongside and assist features of modern electricity markets. For instance, some systems use uplift payments to mitigate market failures~\cite{hogan2003}. The MPMP could serve as an additional tool to collect and deliver uplift payments. Additionally, the MPMP could be used as a collection mechanism to subsidize desirable technologies in the power system (\emph{e.g.}, energy storage, flexible resources, or clean generation) and penalize undesirable ones (\emph{e.g.}, dirty generation). The question of whether a welfare allocation is good or evil is outside the scope of this work.

\section{Conclusion} \label{sec:conclusion}
In this paper, we study the interactions between an energy storage unit (SU) aggregator, its constituent SUs, and the wholesale electricity market.  First, we model the aggregator-SU interactions: we show that long-term cooperative equilibria exist.  Then, we use Nash Bargaining Theory to predict precisely which cooperative equilibrium the aggregator and SUs are likely to settle on. The bargaining outcome has the characteristic of being Pareto-optimal. The Pareto-optimality of the supply/demand bids, allows us to simplify the relationship between the aggregator and its SUs by replacing them with an aggregate profit maximization problem.  Moreover, our results serve as a theoretical justification for instances\footnote{Direct-load control is an example where cooperation between the aggregator and its units is a precondition for the proper functioning of such scheme.} in which aggregators cooperate with their managed units and vice versa.

We also assess the effects of an aggregator on the system welfare.  First, we show that a rational aggregator of SUs is always beneficial to the system.  However, because of market power, the aggregator does not behave in a socially optimal manner.  To remedy this problem, we propose a non-uniform pricing scheme to mitigate the market power of the aggregator.  This pricing scheme has the ability to regulate the profit of the aggregator to a desired amount, its formulation does not require private information, and is compatible with the behavior of a self-interested, profit-seeking aggregator. The interested reader is referred to~\cite{Contreras_PESGM2017} for a case study of the market power mitigating price on the 24 bus IEEE Reliability Test System.

\bibliographystyle{IEEEtran}

\bibliography{bibliography}

\appendix

\subsection*{Proof of  Theorem ~\ref{thm:agg}}
We begin by noting that at the solution of problem~\eqref{eq:syst_prob}, $\bd g_t^*$, $ \zeta_t^*$, $\bd \mu_t^*$, the value of its Lagrangian is equal to the system cost, \emph{i.e.},
\begin{align}
S(\bd q,\bd d) \equiv \mathcal{L}(\bd g_t^*, \zeta_t^*,\bd \mu_t^*, \bd q, \bd d). \label{eq:sys_cost_lag}
\end{align}
From Definition~\ref{defn:LMP}, the system LMPs at time $t$ are defined as
\begin{align*}
\nabla_{\bd q_t} \mathcal{L}= \bd \lambda_t = - \bd 1  \zeta_t - \bd H^\top\bd \mu_t
\end{align*}
and the gradient of the Lagrangian with respect to $\bd d_t$ is
\begin{align} \label{eq:LMP_def}
\nabla_{\bd d_t} \mathcal{L}= -\bd \lambda_t  =  \bd 1  \zeta_t +\bd H^\top\bd \mu_t
\end{align}
both of which are an implicit functions of the load $\bd q$ and storage actions $\bd d$.
\begin{prop}
    The system cost is convex in $\bd d_t$.
\end{prop}
\subsubsection*{Proof}
As shown in~\cite{Xu_Baldick}, the nodal prices are piece-wise functions of the net load $\bd q_t - \bd d_t^\mathrm{bus}$. Each segment of the nodal piece-wise price functions are associated with a set of binding transmission constraints. Within each segment, the nodal price functions are differentiable.  Also, as shown in~\cite{Xu_Baldick} the Jacobian of $\mathcal{L}$ with respect to $\bd d_t$ is negative semi-definite. Since  $\nabla_{\bd q_t} \mathcal{L} = -\nabla_{\bd d_t} \mathcal{L}$, the Jacobian of $\mathcal{L}$ with respect to $\bd d_t$ is positive semi-definite. 
We therefore conclude that the system cost is convex in $\bd d_t$. \QEDB

\begin{prop} \label{prop:thm1_2}
    The net revenue of the aggregator throughout the horizon $\T$ is less than or equal to the change in system cost:
    \begin{align} \label{eq:chg_syst_cost}
    \sum_{t \in \T}\bd \lambda_t^\top \bd d_t^\mathrm{bus} \le \bd S(\bd q, \bd 0) - \bd S(\bd q, \bd d).
    \end{align}
\end{prop}

\subsubsection*{Proof}

From Equations~\eqref{eq:sys_cost_lag} and~\eqref{eq:LMP_def} we can write the left-hand side of equation~\eqref{eq:chg_syst_cost} as
\begin{align} \label{eq:definition_of_lmp}
-\sum_{t \in \T}\nabla_{\bd d_t} S(\bd q, \bd d)^\top\bd d_t^\mathrm{bus}.
\end{align}
Substituting~\eqref{eq:definition_of_lmp} in the left-hand side of Equation~\eqref{eq:chg_syst_cost} and rearranging we get
\begin{align} \label{eq:chg_syst_cost_single_2}
\bd S(\bd q, \bd d)-\sum_{t \in \T}\nabla_{\bd d_t} S(\bd q, \bd d)^\top\bd d_t^\mathrm{bus} \le \bd S(\bd q, \bd 0) .
\end{align}
Note that since $S$ is convex in $\bd d_t$, $\bd S(\bd q, \bd d)-\sum_{t \in \T}\nabla_{\bd d_t} S^\top\bd d_t^\mathrm{bus}$ is an under-estimator of $\bd S(\bd q, \bd 0)$. Thus, the inequality in~\eqref{eq:chg_syst_cost} and~\eqref{eq:chg_syst_cost_single_2} hold. \QEDB

Let $\phi$ denote the non-negative compensation for the SUs under the aggregator. Then, the aggregator's profit can be written as its net revenue minus the compensation to the SUs:
\begin{align*}
\sum_{t \in \T}\bd \lambda_t^\top \bd d_t^\mathrm{bus} - \phi.
\end{align*}
From the non-adversarial behavior of the aggregator, a lower bound for the aggregator's profit is zero. Since $\phi$ is non-negative, a lower bound for the aggregator's net revenue is zero. From Proposition~\ref{prop:thm1_2}, the aggregator's net revenue (zero) is a lower bound of $\bd S(\bd q, \bd 0) - \bd S(\bd q, \bd d)$. Thus, Equation~\eqref{eqn:agg_welfare} holds. This concludes the Proof of Theorem ~\ref{thm:agg}. \QEDB

    \subsection*{Proof of Lemma \ref{lem:coop}}
    The long-term profit made by the aggregator can be written as
    \begin{align}
    \pi^{\mathrm{a},\infty} \!&= \sum_{k=0}^{\infty}{ \delta^k\pi^\mathrm{a}\left(\boldsymbol{\tau}^{(k)}; \boldsymbol{d}^{(k)}\right) } = \sum_{k=0}^{v_i - 1}\!\! \delta^k\pi^\mathrm{a}\!\!\left(\widehat{\boldsymbol{\tau}}; \widehat{\boldsymbol{d}}\right) +\!\!\sum_{k=v_i}^{\infty} \delta^k{\pi_{i}^\mathrm{a}}' \label{eq:longtermaggprofit}
    \end{align}
    where the aggregator cooperates (plays $\widehat{\boldsymbol{\tau}}_i$) until it defects from cooperation with SU $i$ by playing $\boldsymbol{\tau}_i^\mathrm{D}$ at time $v_i$. The aggregator chooses $v_i$ such that its long-term profit is maximized. Both the aggregator and the SU play their defection equilibrium for all subsequent time periods.

    Using the identity $\sum_{k=a+1}^{b-1}{\delta^k}=\frac{\delta^{a+1}-\delta^b}{1-\delta}$, \eqref{eq:longtermaggprofit} can be rewritten as
    \begin{align*}
    \pi^{\mathrm{a},\infty} = \frac{\pi^\mathrm{a}\!\left(\!\widehat{\boldsymbol{\tau}}; \widehat{\boldsymbol{d}}\!\right)}{1-\delta} -\frac{\delta^{v_i}  \!\left(  \pi_{i}^\mathrm{a}\!\left(\!\widehat{\boldsymbol{\tau}}; \widehat{\boldsymbol{d}}\!\right) - {\pi_{i}^\mathrm{a}}' \! \right)}{1-\delta}.
    \end{align*}
    Since $\delta$ is strictly less than 1 and strictly greater than zero $\delta^{v_i}\rightarrow 0$ as $v_i \rightarrow \infty$. It follows that the aggregator maximizes its profit by cooperating indefinitely (\emph{i.e.} chooses $v_i^*=\infty$) if
    \begin{align}
    \pi^\mathrm{a}\!\left(\!\widehat{\boldsymbol{\tau}}; \widehat{\boldsymbol{d}}\!\right)- {\pi_{i}^\mathrm{a}}' \label{eq:longtermaggprofit_2}
    \end{align}
    is greater than zero. We assume that $v_i^*=\infty$ if \eqref{eq:longtermaggprofit_2} is equals to zero. It follows that for the aggregator to cooperate indefinitely with SU $i$, the agreed SU actions and price schedule must deliver a profit greater than ${\pi_{i}^\mathrm{a}}'$.

    Similarly, the SU cooperates indefinitely if
    \begin{align*}
    \pi_{i}^\mathrm{s}\!\left(\!\widehat{\boldsymbol{d}}_i; \widehat{\boldsymbol{\tau}}_i\!\right)-\!(1\!-\!\delta)\pi_{i}^\mathrm{s}(\bd{d}_i^\mathrm{D}; \widehat{\btau}_i)  \!-\!\delta{\pi_{i}^\mathrm{s}}' 
    \end{align*}
    is nonnegative.  It follows that in order for SU $i$ to cooperate indefinitely, the agreed SU actions and price schedule must deliver a profit greater than $(1\!-\!\delta)\pi_{i}^\mathrm{s}(\bd{d}_i^\mathrm{D}; \widehat{\btau}_i)  +\delta{\pi_{i}^\mathrm{s}}'$.\QEDB

\subsection*{Proof of Lemma \ref{lem:bargain}}

 From \cite{Binmore}, the Nash Bargaining Solution is given by
\begin{align}
(\widehat{\btau}_i^*,\widehat{\boldsymbol{d}}_i^*) =\xi\left(\mathcal{B}_i\right) = \argmax_{( \boldsymbol{\tau}_i, \boldsymbol{d}_i) \in \mathcal{B}_i } { \left(\pi_i^\mathrm{s} - {\pi_i^\mathrm{s}}'   \right)\left(\pi^\mathrm{a} - {\pi_i^\mathrm{a}}'  \right) } .\label{eq:xifunction}
\end{align}
By the Pareto-efficiency axiom, the aggregator reaches an agreement with \emph{every} SU and shares the maximum possible profit $\pi(\boldsymbol{d}^*)$ during every single-shot game~\cite{krishna}.

Let $\mathcal{A}_i = \{\!(\boldsymbol{\tau}_i,\boldsymbol{d}_i) | \pi^\mathrm{a}(\btau; \boldsymbol{d} )\!\ge {\pi_{i}^\mathrm{a}}', \pi_{i}^\mathrm{s}(\boldsymbol{d}_i; \boldsymbol{\tau}_i  )\!\ge\!(1\!-\!\delta)\pi_{i}^\mathrm{s}(\bd{d}_i^\mathrm{D}; \btau_i)\!+\! \delta {\pi_{i}^\mathrm{s}}' \}. $By Pareto-efficiency, the agreed prices prices $\widehat{\boldsymbol{\tau}}_i^*$ will be such such that long-term cooperation is sustained, \emph{i.e.} $(\widehat{\boldsymbol{\tau}}_i^*,\widehat{\boldsymbol{d}}_i^*) \in \mathcal{A}_i\; \forall i \in \mathcal{I}$.

Denote the set of $\boldsymbol{\tau}_i$ and $\boldsymbol{d}_i$ such that the maximum profit is split and that long-term cooperation is sustained as $\tilde{\mathcal{B}}_i = \{ (\boldsymbol{d}_i,\boldsymbol{\tau}_i ) | \bd \tau_i \in \mathcal{S}_i^\mathrm{s},\;\boldsymbol{d}_i =  \boldsymbol{d}_i^*,\; (\boldsymbol{d}_i,\boldsymbol{\tau}_i ) \in \mathcal{A}_i  \}$. By the independence of irrelevant alternatives axiom, $\xi\left(\mathcal{B}_i\right) =\xi ( \tilde{\mathcal{B}}_i)$. Then, \eqref{eq:xifunction} can be replaced by
\begin{align*}
\xi\left(\mathcal{B}_i\right) = \argmax_{( \boldsymbol{\tau}_i, \boldsymbol{d}_i) \in \tilde{\mathcal{B}}_i  } { \left(\pi_i^\mathrm{s} - {\pi_i^\mathrm{s}}'  \right)\left(\pi^\mathrm{a} - {\pi_i^\mathrm{a}}'  \right) }
\end{align*}
which is equivalent to problem \eqref{eq:bargprob}.\QEDB

\subsection*{Proof of Theorem \ref{thm:mpmp}}

Using \eqref{eqn:pi_s}, \eqref{eqn:pi_a}, and \eqref{eq:d_bus_defn}, problem~\eqref{eq:Marketclear} can be written as:
\begin{align}\label{eqn:pi_2}
\max_{\bd d \in \mathcal{S}^\mathrm{s}} \sum_{b \in \mathcal{B},\; t \in \T} \lambda_{b,t} \cdot d_{b,t}^\mathrm{bus} - \sum_{i \in \mathcal{I}} c_i(\bd d_i)
\end{align}
where $\lambda_{b,t}$ and $d_{b,t}^\mathrm{bus}$ are the $b^\mathrm{th}$ elements of $\bd \lambda_{t}$ and $\bd d_{t}^\mathrm{bus}$, respectively.

The LMPs $\bd \lambda_t$ are a function of the market clearing process denoted by problem~\eqref{eq:syst_prob}. Thus, problem~\eqref{eq:Marketclear} is implicitly a bi-level problem: the aggregator determines its quantity $\bd d$ followed by the market clearing process which determines the LMPs $\bd \lambda_t$. Denote this problem as \textbf{P1}.

Similarly, the social optimal operation of the SUs defined by problem~\eqref{eq:dictatorprob} is a bi-level problem: the system operator determines the optimal storage actions $\bd d^\mathrm{social}$ by minimizing the system cost $S(\bd q, \bd d)$ which is the result of the market clearing problem~\eqref{eq:syst_prob}. Denote this problem as \textbf{P2}.

Note that \textbf{P1} and \textbf{P2} are similar problems: they have the same set of decision variables, $\bd d$ and $\{\bd g_t\}_{t \in\T}$, and the same set of constraints. Both problems determine the storage actions $\bd d$ (subject to the SU operating constraints) on the upper-level and the generation levels $\bd g_t\;\forall \;t \in \T$ (subject to system constraints) on the lower level. Their objectives, however, differ. The objective of \textbf{P1} is to maximize the aggregator's profit while the objective of \textbf{P2} is to minimize the system cost.

Now we show that replacing the price that the aggregator is exposed from $\lambda_{b,t}$ to $p_{b,t}$ equates the objectives of \textbf{P1} and \textbf{P2}. Replacing $\lambda_{b,t}$ with the definition of $p_{b,t}$ from Theorem~\ref{thm:mpmp} in~\eqref{eqn:pi_2} we get
\begin{align*}
\max_{\bd d \in \mathcal{S}^\mathrm{s}} \sum_{b \in \mathcal{B}, t \in \T}-\int_0^{g_{b,t}} \!\!\!c_{b,t}^\mathrm{gen}(x)dx - \sum_{i \in \mathcal{I}} c_i(\bd d_i) +\sum_{b \in \mathcal{B}} C_{b,t},
\end{align*}
which is equivalent to the system cost minimization objective in Equation~\eqref{eqn:Sys_obj} plus a constant term $\sum_{b \in \mathcal{B}} C_{b,t}.$

We have shown that the objectives of \textbf{P1} and \textbf{P2} are equivalent when the aggregator is exposed to the MPMP $p_{b,t}$. Furthermore, we showed that \textbf{P1} and \textbf{P2} are subject to the same set of constraints. Thus we conclude that when the aggregator is exposed to $p_{b,t}$ rather than to the LMP $\lambda_{b,t}$, \textbf{P1} and \textbf{P2} are equivalent problems. Thus, the market clears at the social optimum when the aggregator is exposed to $p_{b,t}$. \QEDB

\end{document}